\documentclass[12pt,preprint]{aastex}

\def\gtorder{\mathrel{\raise.3ex\hbox{$>$}\mkern-14mu
             \lower0.6ex\hbox{$\sim$}}}
\newcommand{\chandra}{\textit{Chandra}}
\newcommand{\chandralong}{\textit{Chandra X-ray Observatory}}

\newcommand{\xray}{\mbox{X-ray}}

\def\ltsima{$\; \buildrel < \over \sim \;$}
\def\simlt{\lower.5ex\hbox{\ltsima}}
\def\gtsima{$\; \buildrel > \over \sim \;$}
\def\simgt{\lower.5ex\hbox{\gtsima}}

\begin{document}


\title{The $J$-band Light Curve of SN 2003lw, Associated
with GRB 031203}


\author{A. Gal-Yam\altaffilmark{1,2}, D.-S. Moon\altaffilmark{1,3}, D. B. Fox\altaffilmark{1}, A. M. Soderberg\altaffilmark{1}, S. R. Kulkarni\altaffilmark{1}, E. Berger\altaffilmark{1},
S. B. Cenko\altaffilmark{1}, S. Yost\altaffilmark{1},
D. A. Frail\altaffilmark{4}, 
M. Sako\altaffilmark{5},
W. L. Freedman\altaffilmark{6}, S. E. Persson\altaffilmark{6}, P. Wyatt\altaffilmark{6}, D. C. Murphy\altaffilmark{6},
M. M. Phillips\altaffilmark{7},
N. B. Suntzeff\altaffilmark{8},
P. A. Mazzali\altaffilmark{9,10}
K. Nomoto\altaffilmark{10}
}
\email{avishay@astro.caltech.edu}

\altaffiltext{1}{Division of Physics, Mathematics, \& Astronomy, MS 105-24, California Institute of Technology,
Pasadena, CA 91125}
\altaffiltext{2}{Hubble Fellow}
\altaffiltext{3}{Robert A. Millikan Fellow}
\altaffiltext{4}{National Radio Astronomy Observatory, P.O. Box 0, Socorro, New Mexico 87801}
\altaffiltext{5}{Stanford Linear Accelerator Center, 2575 Sand Hill Road M/S 29, Menlo Park, CA 94025}
\altaffiltext{6}{Observatories of the Carnegie Institution of Washington, 813 Santa Barbara Street, Pasadena, CA 91101}
\altaffiltext{7}{Las Campanas Observatory, Carnegie Observatories, Casilla 601, La Serena, Chile}
\altaffiltext{8}{Cerro Tololo Inter-American Observatory, National Optical Astronomy Observatory, Casilla 603, La Serena, Chile}
\altaffiltext{9}{INAF, Osservatorio Astronomico di Trieste, Via Tiepolo, 11, I-34131 Trieste, Italy}
\altaffiltext{10}{Dept. of Astronomy and RESCEU, University of Tokyo, Tokyo 113-0033, Japan}

\begin{abstract}

At $z=0.1055$, the gamma-ray burst GRB 031203 is one of the two nearest GRBs known. 
Using observations from the Very Large Array (VLA) and \chandralong\, 
we derive sub-arcsecond localizations of the radio and
X-ray afterglow of this GRB. We present near-infrared observations of 
the supernova SN 2003lw, which exploded in the
host galaxy of the GRB 031203. Our deep, high resolution Magellan/PANIC 
data establish that this SN is spatially coincident with the
radio and X-ray localizations of the afterglow of GRB 031203 to sub-arcsecond 
precision, and is thus firmly associated with the GRB. 
We use image differencing to subtract the bright
emission from the host galaxy, and measure the time evolution of the SN
between $\sim5$ and $\sim50$ days after the GRB.   
The resulting light curve has a shape which is quite different from 
that of the two SNe previously associated with GRBs, SN 1998bw and SN 2003dh.  
With SN 2003lw securely associated with this burst, we confirm that all three GRBs with 
redshifts $z<0.3$ were accompanied by SN explosions. 

\end{abstract}




\section{Introduction}

The emerging association between long-duration gamma-ray bursts (GRBs)
and type Ic supernovae (SNe Ic) is perhaps the most significant breakthrough
in our understanding of GRBs, and may also provide new insights into
the physics of core-collapse SNe and the deaths of massive stars.
The initial strong evidence for this connection came from the spatial
and temporal coincidence between GRB 980425 and SN 1998bw (Galama et al. 1998).
More recently, this picture was convincingly affirmed by the detection
of SN features in the optical afterglow spectrum of GRB 030329 (Stanek et al. 2003;
Hjorth et al. 2003; Matheson et al. 2003). Both of these milestone discoveries
resulted from the study of the nearest GRBs yet identified, GRB 980425 
at $z=0.0085$ and GRB 030329 at $z=0.1685$. Indeed, each relatively rare occurrence of a GRB
at low redshift ($z<0.3$) provides a unique opportunity for further study of 
the GRB-SN connection.
  
GRB 031203\footnotemark~was detected by IBIS on board the INTEGRAL spacecraft on 
2003 December 3, at 22:01:28 UT (Gotz et al. 2003). 
A fading X-ray afterglow was discovered by  XMM-Newton on 2003 December 4 
(Santos-Lleo et al. 2003; Campana et al. 2003; Rodriguez-Pascual et al. 2003) 
and found to be consistent with the location of a radio transient (Frail 2003; Soderberg,
Kulkarni, \& Frail 2003). Prochaska et al. (2003a; 2003b; 2004) identified the
host galaxy of the X-ray and radio transients, determined its redshift ($z=0.1055$)
and studied its properties in detail. Recently, several groups (Bersier et al. 2004; 
Tagliaferri et al. 2004a; Thomsen et al. 2004; Cobb et al. 2004) reported optical 
photometric and spectroscopic observations of 
this GRB, which apparently reveal the signatures of an associated SN (designated 
SN 2003lw; Tagliaferri et al. 2004b). In spite of the 
low redshift of this event, its low Galactic latitude (less than $5^{\circ}$ from
the plane) and the resulting Galactic extinction ($E[B-V]=1.04$; 
Schlegel, Finkbeiner, \& Davis 1998),
as well as its bright host galaxy, make the study of this event challenging.

\footnotetext{The possible classification of this event as an X-ray flash (XRF) was debated
in recent literature (e.g., Prochaska et al. 2004; Watson et al. 2004; Thomsen et al. 2004).
However, the high-energy observations presented by Sazonov, Lutovinov, \& Sunyaev (2004)
conclusively show that this event does not fit any of the commonly used definitions of
XRFs.}

To overcome these difficulties, we have undertaken near infrared (NIR) observations.
The results of this effort are reported in this {\it Letter}.
Coordinated radio and X-ray observations, which enabled us to probe the total 
energy output of this sub-energetic GRB, are reported elsewhere (Soderberg et al. 2004).

\section{Observations}

\subsection{Near Infrared imaging}

The location of the X-ray and radio transients associated with
GRB 031203 was observed with Persson's Auxiliary Nasmyth Infrared Camera 
(PANIC; equipped with a $1024 \times 1024$ HgCdTe Hawaii I infrared focal plane array), 
mounted on the Baade (Magellan I) 6.5m telescope, during 5 nights, in
the $J$ ($1.25\mu$) and $K_s$ ($2.16\mu$) bands. Due to their lower sensitivity and poor
temporal coverage, we did not detect SN 2003lw in our $K_s$-band 
data. We therefore report in this {\it Letter} only the analysis
of our $J$-band observations. We obtained 18 or 36 dithered frames 
during four epochs, approximately 5, 7, 50, and 81 days after 
the GRB. A journal of the observations is given in Table 1. 
The integration time for each frame was 120 s. 
We obtained a common sky frame for each data set from the stacked image cube, and subtracted
it from each frame. We then carried out flat fielding, and shifted each frame
to a common reference position to create the final image.

\begin{deluxetable}{llll}
\tablecolumns{4}
\tablewidth{0pt}
\tablecaption{$J$-band observations of SN 2003lw\label{TableObservations}}
\tablehead{
\colhead{Epoch} & \colhead{UT date} & \colhead{Days after} & \colhead{Number of}\\
\colhead{} & \colhead{} & \colhead{GRB 031203} & \colhead{120s frames}
}
\startdata
 $01$  & 2003 December 9,  07:07:39  &   $5.38$       & $18$                       \\
 $02$  & 2003 December 11, 07:20:05  &   $7.39$       & $18$                       \\
 $03$  & 2004 January  23, 03:10:01  &  $50.21$       & $36\times2$\tablenotemark{a}\\
 $04$  & 2004 February 23, 00:45:25  &  $81.11$       & $36$                       \\
 \enddata
\tablenotetext{a}{Data from two consecutive nights were combined.}
\end{deluxetable}

\subsection{Image subtraction and SN photometry}

In order to extract the faint SN signal from the bright host galaxy background,
we have used the Common Point-spread-function (PSF) image subtraction method 
(CPM; Gal-Yam, Poznanski, \& Maoz, in preparation)\footnotemark. We find that
extra flux is detected in our three early epochs (obtained 5.38, 7.39 and 50.21 days
after the GRB) when compared
with our latest epoch (day 81.11; see Fig. 1). We therefore proceed in our analysis by assuming
that our latest image contains only a negligible amount of SN light. This is supported
by $J$-band observations presented by Cobb et al. (2004). 

\footnotetext{CPM compares a new image with a reference one by extracting
an empirical point-spread-function (PSF) from both images, and convolving each
image with the PSF of the other. While slightly degrading the final seeing
(as the final PSF of both images is worse than the initial PSF of either image), this
procedure produces output images with nominally identical PSFs, while introducing 
a minimal amount of noise. In particular, we have found that this algorithm often
produces difference images which have lower subtraction residuals near the nuclei of
bright galaxies compared to other popular image subtraction packages, and thus allows
better sensitivity to variable sources which are superposed on bright galactic background.} 

To accurately measure the SN flux in each of the difference frames (Fig. 1, panels a-c),
we have added an artificial point source, of known negative flux and with the native PSF,
to our reference image. This results in the introduction of a positive artificial star
to each of the difference images, which should have an identical PSF to any real 
variable point source (the native PSF of the reference image convolved with the PSF
of the image from which it was subtracted). This source is visible in panels a-c of
Fig. 1, in the lower left quadrant. 
Using aperture photometry (with the aperture equal to the FWHM of the artificial
star), we measure the flux of SN 2003lw with respect to this constant source,
and derive the relative photometry of SN 2003lw. We have identified 22 objects which 
appear in the 2MASS point source catalog and fall into the PANIC field of view. Of these,
the brightest 6 are saturated or non-linear in our images. We have used the 
remaining 16 objects to tie our zero-point to the 2MASS photometry. We then
used 10 secondary calibrators in the close vicinity of GRB 031203 to set the 
absolute calibration of the artificial reference star.
From the scatter in calibrations derived using different secondary calibrators,
we estimate a final absolute calibration error of $0.08$ mag. 
Assuming this zero point, we calculate the $J$-band magnitudes of SN 2003lw at 5.38, 
7.39 and 50.21 days after GRB 031203, which are reported in Table 2 and plotted in Fig. 3. 
Our calibration yields $J$-band magnitudes for the 
host galaxy and the SN which are consistent with those reported by Cobb et al. (2004).

\begin{deluxetable}{lll}
\tablecolumns{3}
\tablewidth{0pt}
\tablecaption{$J$-band photometry of SN 2003lw\label{TableObservations}}
\tablehead{
\colhead{UT date} & \colhead{Days after} & \colhead{$J$ magnitude} \\
\colhead{} & \colhead {GRB 031203} & \colhead{}
}
\startdata
 2003 December 9,  07:07:39  &   $5.38$       & $21.66\pm0.60$                       \\
 2003 December 11, 07:20:05  &   $7.39$       & $20.57\pm0.21$                       \\
 2004 January  23, 03:10:01  &  $50.21$       & $20.60\pm0.21$                \\
 \enddata
\end{deluxetable}

The main source of error in the derivation of the magnitudes reported here is the
uneven background levels of the difference images, caused by effects such as non-perfect
flat-fielding and sky subtraction (as seen in panels a-c
of Fig. 1). In order to quantify this error, we have added 20 artificial point sources
to each of the images, at random positions sampling both empty sky and high-background
areas of the image, with the appropriate PSF. We then subtracted the reference image using
CPM. We photometered the resulting difference frames and have iteratively adjusted the flux of the 
artificial sources until their mean measured flux from the difference image was equal to
the flux we measured for SN 2003lw at the same epoch. Finally, we adopted the measured scatter
in the magnitudes of the artificial sources as the error in the SN magnitude at that epoch. 
We find that these errors are far larger than other sources of error, e.g., 
the Poisson errors in the aperture photometry 
of the bright artificial reference star we used in our photometric sequence, which are therefore
neglected. To these errors we add in quadrature our absolute calibration uncertainty 
given above.

\begin{figure*}
\includegraphics[width=17cm]{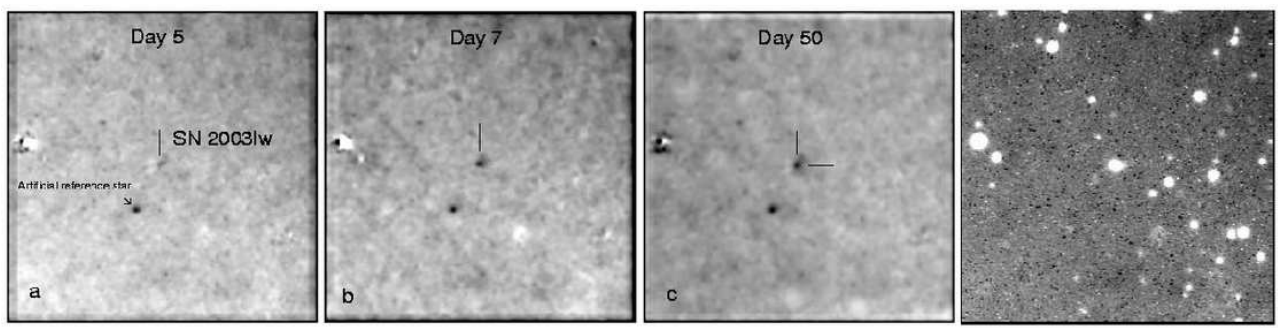}
\caption{NIR detection of SN 2003lw. Panels a-c show $37.7"\times37.7"$ sections of
the difference images obtained by subtracting our last observed epoch (81 
days after the GRB) from the first three epochs, obtained at
5.38, 7.39 and 50.21 days post burst. SN 2003lw is at the center of each frame
(indicated by crosshairs) and its brightness can be compared to an artificial
reference star added to each image (marked in panel a) which has a constant flux.
The Grey levels in panels a-c are linearly scaled between the peak of the 
artificial reference star and $1\sigma$ of the sky noise below the mode sky level. 
Comparing panels b and c,
note the similar IR flux detected at 7 and 50 days after the GRB. For orientation
purposes, panel d shows a section from our reference (fourth epoch) PANIC image with the
same scale and orientation. The host galaxy of GRB 031203/SN 2003lw lies at the
center of panel d. North is up, East to the left.}   
\end{figure*} 

\section{Results}

\subsection{Spatial Coincidence with the Radio and X-ray Afterglow of
            GRB\,031203}

The position of the radio afterglow from VLA observations (Soderberg et al.
2004) is (J2000) RA 08:02:30.1833(18), Dec $-$39:51:03.522(78), as
referenced to the ICRS (uncertainties in the final digits are given in
parentheses). This position is derived from our first-epoch 22~GHz
observation, the highest resolution of our various radio observations,
which had a beam size of 1.14$\times$0.29 arcsec.

The position of the \xray\ afterglow is derived from our \chandralong\
observations.  We observed the afterglow with \chandra\ in a single
21.6~ksec exposure beginning at 21:35~UT on 22~January 2004 (mean
epoch 49.1~days post-burst), with the afterglow position at the
aimpoint of the \mbox{ACIS-S3} CCD.  Data were reduced in the standard
manner using the CIAO tools (v.~3.0.2).  In particular, a
comprehensive ``wavdetect'' source detection analysis reveals six
X-ray sources on the S3 chip with coincident objects in the
\mbox{GSC-2.2} catalog.  Using these sources we confirm the absolute
astrometry to $\pm$0.13 arcsec precision, and derive a position for
the afterglow of (J2000) RA 08:02:30.159, Dec $-$39:51:03.51, with
0.18\arcsec\ uncertainty.  We then identify two X-ray sources on the
S3 chip with point-like sources in our 9~December 2003 PANIC $J$-band
image.  By registering these two \xray\ source positions against their
$J$-band counterparts, we are able to locate the afterglow on the
PANIC image to $\approx$1.2-pixel precision.

The resulting X-ray localization,
0.56$\times$0.64 arcsec in size, is shown as the red ellipse in
Figure~2, along with the VLA radio (cyan) and PANIC NIR
transient (green) localizations, as transferred to the same image.
All ellipses are two-sigma ($95\%$ confidence level).  
The WCS on the PANIC image is derived
from the positions of 23 2MASS counterparts, and the uncertainty in
the VLA position is dominated by the 0.095$\times$0.121 arcsec RMS of
this mapping.  The NIR localization is derived from two of our
subtracted images; the two relative localizations differ by (0.23,
0.46) pixels and so we adopt 0.029$\times$0.058 arcsec as our
one-sigma uncertainty.  

Fig.~2 shows that the location of the NIR transient is
consistent with that of the afterglow across all wavelengths, to high
precision, confirming the association that has already been inferred
from the temporal coincidence of GRB\,031203 and SN\,2003lw.

Moreover, this location is consistent with the peak of the host galaxy
light: the NIR offset, which has the highest precision, is
0.15$\pm$0.23~pixels West, 0.26$\pm$0.46~pixels North.  At $z=0.1055$,
this 0.04$\pm$0.06 arcsec offset corresponds to a physical distance of
0.07$\pm$0.13~kpc.


\begin{figure}
\label{fig:panic}
\includegraphics[width=17cm]{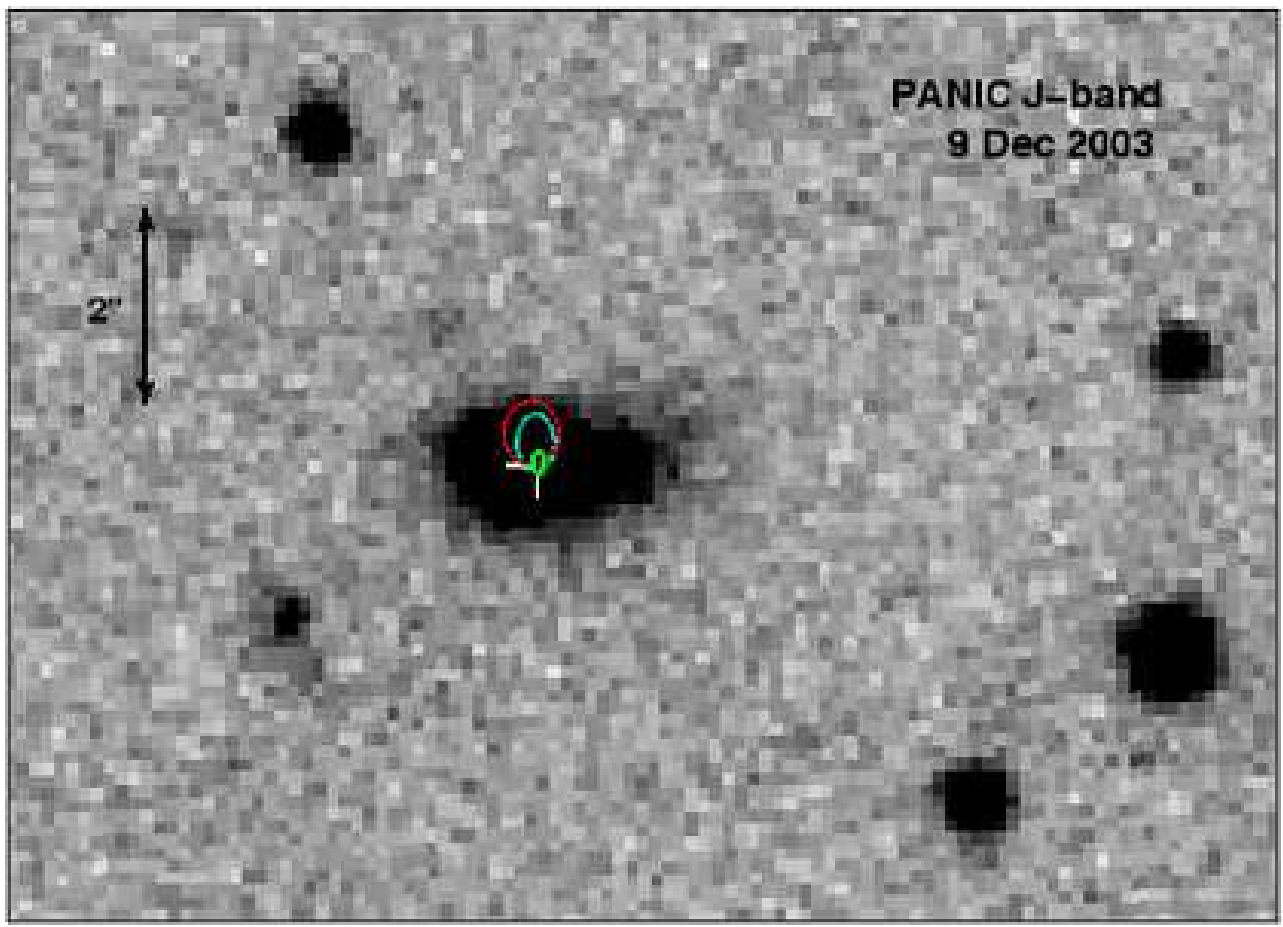}
\caption{
PANIC $J$-band image of the host galaxy of GRB\,031203 from 9~December
2003, with localizations of the \chandra\ \xray\ afterglow (red), the
VLA radio afterglow (cyan), and the PANIC NIR transient (green)
superposed; the position of the central maximum of the host galaxy
light is indicated by the white ticks, and North is up and East is to
the left.  The ellipses shown are two-sigma (95\%-confidence)
localizations; see text for details.  The NIR transient ellipse is
0.46$\times$0.92 pixels (0.06$\times$0.11 arcsec) in size.  The size
of the VLA ellipse is 0.38$\times$0.58 arcsec; the X-ray ellipse is
0.56$\times$0.64 arcsec.  Although transient light contributes to this
image, its effect on the appearance of the host galaxy at this color
stretch is negligible.  The localizations of the afterglow and
transient are consistent across all wavelengths, and locate the source
within 0.2~kpc of the central maximum of the host galaxy light.
}   
\end{figure} 

\subsection{The J-band Light Curve}

Figure 3 shows the $J$-band light curve of SN 2003lw. Our deep observations detect
the SN light as early as $\sim5$ days after the burst (the earliest detection of 
this SN), and suggest a steep rise in flux
between the first two epochs ($\sim5$ and $\sim7$ days after the GRB). The flux 
measurements at $\sim7$ and $\sim50$ days after the GRB are comparable.

To better interpret these observations,
we wish to compare our data to the $J$-band light curve of SN 1998bw, the prototype
and best observed GRB-associated SN. Unfortunately, NIR observations of SN 1998bw
are scarce - the only data available are 3 epochs of photometry and spectroscopy 
reported by Patat et al. (2001). We have used these NIR spectra, in conjunction with
optical spectroscopy from the same source, to calculate the $I-J$
colors of SN 1998bw, redshifted to $z=0.1055$, using the methods described
by Poznanski et al. (2002). We then construct a
model for the $J$-band light curve of SN 1998bw as it would appear at $z=0.1055$
in the following manner. Based on the observations of Galama et al. (1998) we construct 
the I-band light curve of SN 1998bw at $z=0.1055$, using the extinction
correction given by Cobb et al. (2004). The resulting curve is shown in Fig. 3 
(dashed curve). We then use the three epochs of $I-J$ colors calculated above, along with
the appropriate (smaller) $J$-band extinction correction (from Prochaska
et al. 2004) to calculate $J$-band anchors for our light curve model (asterisks in Fig. 3). Finally, we assume
that the shapes of the $J$ and $I$-band light curves of SN 1998bw are similar, and
adjust the $I$-band light curve upward by the mean of the three color measurements we have,
yielding the solid curve plotted in Fig. 3. Note that using the $I$-band light curve shape we can
obtain a decent fit to all three $J$-band anchors, and thus it appears that assuming
this light curve shape is reasonable, at least during the declining
part of the light curve, sampled by the available NIR photometry of SN 1998bw.  

Having exhausted the relevant observational data, we turn to models of SN 1998bw. 
Iwamoto et al. (1998) present hydrodynamical models of exploding C+O stars
which reproduce the optical light curve and spectra of SN 1998bw. Synthetic 
spectra calculated by these authors extend out to $1.2\mu$. We use these spectra to 
calculate synthetic $I-J$ colors of SN 1998bw (as it would seem at $z=0.1055$) 
between $\sim10$ and $\sim107$ days after the GRB, as we have done with the real
date. Errors resulting from incomplete coverage of the
$J$-band by the available spectra, are calculated as in Poznanski 
et al. (2002). We then repeat the process described above, creating a new set of
synthetic anchors (marked by small filled circles in Fig. 3). The distribution 
of these anchor points suggests again that the $I$ and $J$-band light curves of SN 1998bw
had similar shapes. Indeed, applying the average synthetic color correction $I-J\sim0.9$
and the appropriate extinction correction to our model $I$-band light curve of SN 1998bw, 
the resulting model $J$-band light curve of SN 1998bw (light solid curve in Fig. 3) 
fits the synthetic anchors very well. The $J$ band model light curves derived from
the sparse NIR observations (boldsolid curve) and the 
model spectra (light solid curve) are reasonably consistent with each
other and support the notion that the $I$ and $J$-band light curves of SN 1998bw had similar shape. 
Interestingly, the rough $I-J$ colors of SN 2003lw
measured by Cobb et al. (2004; $I-J\sim1$) appear similar to the colors we derive for SN 1998bw
at $z=0.1055$.

Comparing our model SN 1998bw light curves with the data, an obvious discrepancy is
immediately revealed. While our late-time point is marginally consistent with the models, 
the early-time $J$-band data points fall significantly below model
expectations (by at least $\sim0.9$ mag, a $9\sigma$ effect). 
This discrepancy is enhanced if SN 2003lw was intrinsically brighter than SN 1998bw 
by $\sim0.4$ mag, as advocated by Thomsen et al. (2004). We also note that the fast 
early rise implied by our early data points is not reproduced by either model. 
Making SN 2003lw less luminous overall would make the model inconsistent with
our late time data point, and would conflict with the observations of
Thomsen et al. (2004) and Cobb et al. (2004). 
Thus, it appears that SN 2003lw had a significantly
different light curve shape than SN 1998bw, at least in the $J$-band. 

\begin{figure*}
\includegraphics[width=17cm]{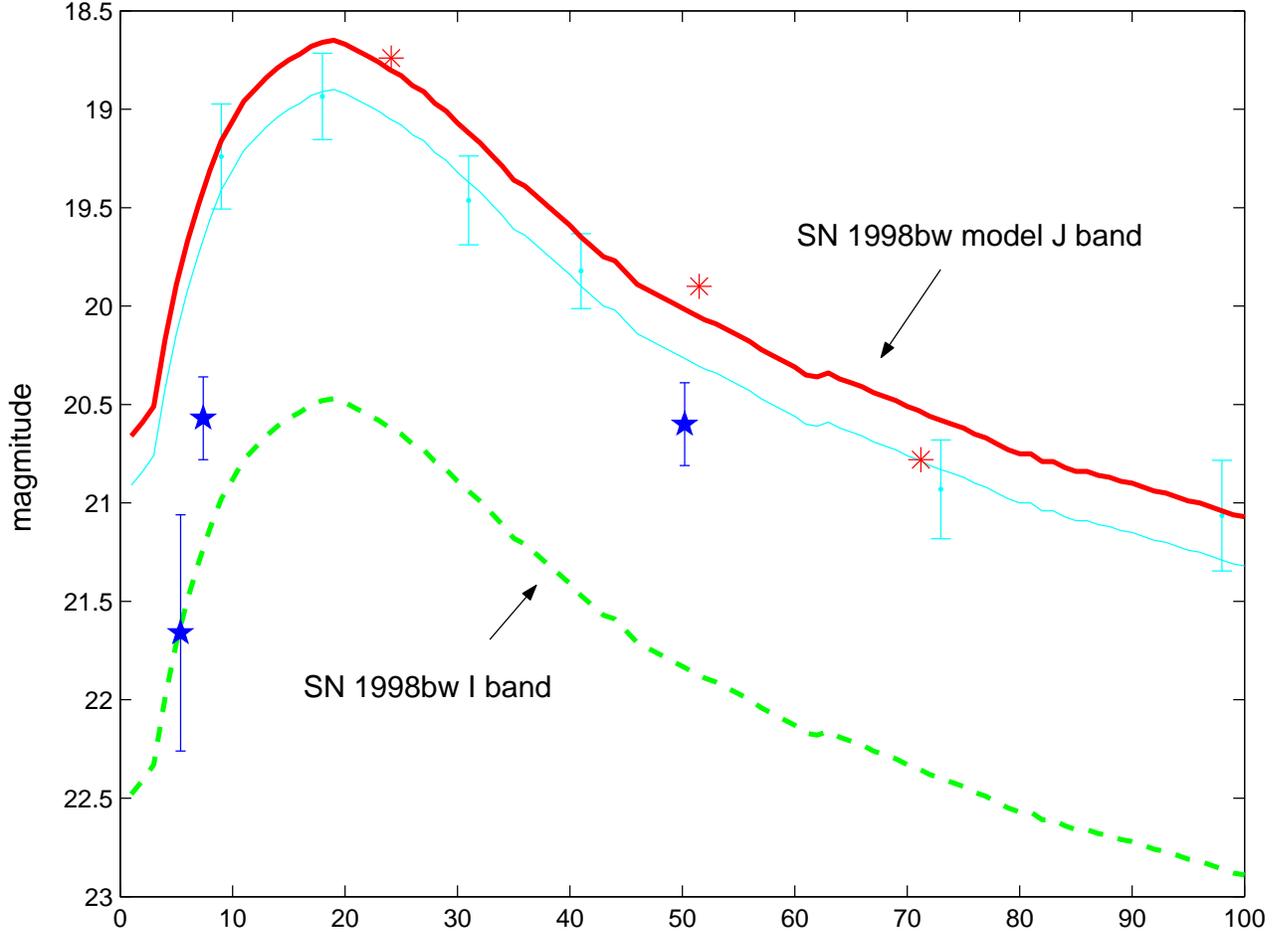}
\caption{$J$-band photometry of SN 2003lw. Our measurements (Stars with
$2\sigma$ error bars, see text) are compared to model light curves
of SN 1998bw. The bold dashed line shows the $I$-band light curve of SN 1998bw,
redshifted to $z=0.1055$, assuming an extinction of $A_I=1.4$ mag (Cobb et al. 2004).
The only available $J$-band
spectra of SN 1998bw (Patat et al. 2001) are used to calculate $I-J$
color offsets, correcting for the different extinction values in the $I$ 
and $J$ bands (from Prochaska et al. 2004; see text). 
These points are plotted as asterisks, and serve as our guidance in constructing
a model $J$-band light curve for SN 1998bw. This is done by adopting the I-band 
light curve shape, normalized to fit these three points (bold solid curve). 
A second model, based on $J$-band points calculated using synthetic color corrections
(small circles, see text) is shown as a light solid curve. Note that both $J$-band 
models are inconsistent with our early data points.
}
\end{figure*} 

\section{Discussion and Conclusions}

In this {\it Letter}, we have presented $J$-band observations of SN 2003lw.
The sensitivity and resolution of our data enabled us to pinpoint
the location of SN 2003lw within its host galaxy, and to show
it is consistent with sub-arcsecond localizations of the
radio and X-ray afterglow of GRB 031203, thus confirming the association
of these two events. The precise NIR localization of this event also
puts it within $0.2$~kpc from the host galaxy center.
The $J$-band light curve of SN 2003lw shows a rapid
initial rise ($5-7$ days after the GRB) 
and evidence for bright emission more than $50$ days after the GRB.
The fast early rise of SN 2003dh, associated with GRB 030329, has been 
interpreted by Woosley \& Heger (2003) and Mazzali et al. (2003) 
as evidence for asymmetry in the explosion.
A thorough investigation of this possibility will
probably require an analysis of our data in combination with other extensive
data sets of optical and NIR photometry and spectroscopy collected by
other groups (e.g., Thomsen et al. 2004; Tagliaferri et al. 2004; Bersier et al. 2004;
Cobb et al. 2004).
  
Cobb et al (2004) have recently reported $I$ and $J$-band observations of 
this event, obtained with the SMARTS 1.3m telescope. A direct comparison between
our observations and the SMARTS data, obtained on numerous epochs, is complicated 
by the fact that these authors do not present the light curve of SN 2003lw. Instead, they
plot the temporal evolution of the combined light
of the SN and its bright host galaxy, derived from aperture photometry,
which shows considerable scatter. It is thus
hard to say whether their data show the same early fast rise we detect. 
Our observation that SN 2003lw had similar
flux levels 7 and 50 days after the GRB is consistent with the reported SMARTS
data. Furthermore, comparing their $I$-band data with model light curves
of SN 1998bw, these authors arrive at the conclusion that 
the light curve shape of SN 2003lw does not resemble that of SN 1998bw. This
is in accord with our analysis of the $J$-band data (Cobb et al. 2004 do not attempt
to compare their $J$-band data with a model of SN 1998bw). It appears that both 
our observations and the Cobb et al. (2004) data set suggest that SN 2003lw had 
a light curve quite unlike that of SN 1998bw; with a fast rise to maximum, which appears 
broader, and perhaps showing a secondary peak in the IR.  

The above discussion provides further evidence for the 
diversity of SNe associated with GRBs [see, e.g., Thomsen et al. (2004) and 
Lipkin et al. (2004) for recent reviews]. 
With SN 2003lw, all three nearby GRBs are firmly associated with SNe,
apparently supporting suggestions made, e.g., by Podsiadlowski et al (2004), 
that all long GRBs are accompanied by SNe.
It may well be the case that the focus of future studies should now move from
proving the association between SNe and GRBs to an attempt to characterize the 
properties of this population of SNe. Such studies may provide new clues about
the progenitors and engines of GRBs, by requiring viable GRB models to be
able to produce the large quantities of nickel derived from the SN observations, 
as well as valuable insights into possible explosion mechanisms of 
core-collapse SNe, which may involve GRB-like aspherical effects [e.g., Khokhlov et al. (1999), 
see Gal-Yam et al. (2004) for further discussion]. Low-redshift GRBs, expected to 
be localized in larger numbers by the upcoming {\it SWIFT} mission, as well as 
systematic studies of local core-collapse SNe (e.g., Berger et al. 2003; Soderberg
et al. 2004, in preparation; Stockdale et al. 2004) will probably shed 
more light on these intriguing questions.

\section*{Acknowledgments}

We thank D. Poznanski for his help with synthetic photometry.
A.G. acknowledges support by NASA through Hubble Fellowship grant
\#HST-HF-01158.01-A awarded by STScI, which is operated by AURA, Inc.,
for NASA, under contract NAS 5-26555. 
The Guide Star Catalog was produced at the Space Telescope Science
Institute under U.S. Government grant. These data are based on
photographic data obtained using the Oschin Schmidt Telescope on
Palomar Mountain and the UK Schmidt Telescope.

\clearpage



\begin{thebibliography}{}

\bibitem{} Berger, E., Kulkarni, S.~R., Frail, D.~A., \& Soderberg, A.~M.\ 2003, \apj, 
599, 408 
\bibitem{} Bersier, D., et al.\ 2004, GRB Circular Network, 2544, 1 
\bibitem{} Campana, S., Tagliaferri, G., Chincarini, G., Covino, S., Fugazza, D., \& Stella, L.\ 
2003, GRB Circular Network, 2478, 1
\bibitem{} Cobb, B.~E., Bailyn, C.~D., van Dokkum, P.~G., Buxton, M.~M., \& 
Bloom, J.~S.\ 2004, ApJL, submitted, ArXiv Astrophysics e-prints, astro-ph/0403510
\bibitem{} Frail, D.~A.\ 2003, GRB Circular Network, 2473, 1 
\bibitem{} Galama, T.~J.,~et al.\ 1998, \nat, 395, 670
\bibitem{} Gal-Yam, A., Poznanski, D., Maoz, D., Filippenko, A.~V., \& Foley, R.~J.\ 2004, PASP,
submitted, ArXiv Astrophysics e-prints, astro-ph/0403296
\bibitem{} Gotz, D., Mereghetti, S., Beck, M., Borkowski, J., \& Mowlavi, N.\ 2003, 
GRB Circular Network, 2459, 1 
\bibitem{} Hjorth, J., et al.\ 2003, \nat, 423, 847 
\bibitem{} Iwamoto, K., et al.\ 1998, \nat, 395, 672
\bibitem{} Khokhlov, A.~M., H{\"o}flich, P.~A., Oran, E.~S., Wheeler, J.~C., Wang, L., \& Chtchelkanova, 
A.~Y.\ 1999, \apjl, 524, L107
\bibitem{} Lipkin, Y.~M.~et al.\ 2003, ApJ, in press, ArXiv Astrophysics e-prints, astro-ph/0312594 
\bibitem{} Matheson, T.,~et al.\ 2003, \apj, 599, 394 
\bibitem{} Mazzali, P.~A., et al.\ 2003, \apjl, 599, L95
\bibitem{} Patat, F.,~et al.\ 2001, \apj, 555, 900 
\bibitem{} Podsiadlowski, P., Mazzali, P.~A., Nomoto, K., Lazzati, D., \& Cappellaro, E.\ 2004, 
ApJL, submitted, ArXiv Astrophysics e-prints, astro-ph/0403399 
\bibitem{} Poznanski, D., Gal-Yam, A., Maoz, D., Filippenko, A.~V., Leonard, D.~C., \& Matheson, T.\ 
2002, \pasp, 114, 833 
\bibitem{} Prochaska, J.~X., Chen, H.~W., Hurley, K., Bloom, J.~S., Graham, J.~R., \& Vacca, W.~D.\ 
2003a, GRB Circular Network, 2475, 1 
\bibitem{} Prochaska, J.~X., Bloom, J.~S., Chen, H.~W., Hurley, K., Dressler, A., \& Osip, D.\ 2003b, GRB 
Circular Network, 2482, 1
\bibitem{} Prochaska, J.~X., et al.\ 2004, ApJ, submitted, ArXiv Astrophysics e-prints, astro-ph/0402085
\bibitem{} Rodriguez-Pascual, P., Santos-Lleo, M., Gonzalez-Riestra, R., Schartel, N., 
\& Altieri, B.\ 2003, GRB Circular Network, 2477, 1
\bibitem{} Sazonov, S. Yu., Lutovinov, A. A., \& Sunyaev, R. A., in prep.
\bibitem{} Santos-Lleo, M., Calderon, P., \& Gotz, D.\ 2003, GRB Circular Network, 2464, 1
\bibitem{} Schlegel, D.~J., Finkbeiner, D.~P., \& Davis, M.\ 1998, \apj, 500, 525
\bibitem{} Soderberg, A.~M., Kulkarni, S.~R., \& Frail, D.~A.\ 2003, GRB Circular 
Network, 2483, 1
\bibitem{} Soderberg, A. M., et al. 2004, in prep.
\bibitem{} Stanek, K.~Z.,~et al.\ 2003, \apjl, 591, L17 
\bibitem{} Stockdale, C.~J., Van Dyk, S.~D., Sramek, R.~A., Weiler, K.~W., Panagia, N., Rupen, M.~P., \& 
Paczynski, B.\ 2004, \iaucirc, 8282, 2 
\bibitem{} Tagliaferri, G., et al.\ 2004a, GRB Circular Network, 2545, 1
\bibitem{} Tagliaferri, G., et al.\ 2004b, IAU Circular 8303
\bibitem{} Thomsen, B., et al.\ 2004, A\&A, submitted, ArXiv Astrophysics e-prints, astro-ph/0403451
\bibitem{} Watson, D., et al.\ 2004, ApJL, submitted, ArXiv Astrophysics e-prints, astro-ph/0401225
\bibitem{} Woosley, S.~E.~\& Heger, A.\ 2003, ApJ, submitted, ArXiv Astrophysics e-prints, astro-ph/0309165 

\end{thebibliography}
\end{document}